\documentclass[12pt]{article}
\usepackage{amsmath}
\usepackage{amsfonts,amsthm}
\usepackage{latexsym}
\usepackage{amscd}
\usepackage{amssymb}
\usepackage{graphicx}
\usepackage{float}
\usepackage[applemac]{inputenc}
\usepackage{dsfont}
\usepackage{multirow}
\usepackage{dcolumn}
\usepackage{rccol}
\usepackage{enumerate}
\usepackage{scalefnt}
\usepackage{booktabs}
\usepackage{subfigure}
\usepackage{geometry} 
 \usepackage[table]{xcolor}
\usepackage{longtable}
\usepackage{multirow}
\usepackage{multicol}
\usepackage{lineno}
\usepackage[pdftex,colorlinks=true]{hyperref}
\hypersetup{colorlinks=true, linkcolor=red, filecolor=blue, pagecolor=blue, urlcolor=green, citecolor=blue}

\usepackage{textcomp}
\usepackage{array}
\usepackage{supertabular}
\usepackage{hhline}

\makeatletter
\newcommand\arraybslash{\let\\\@arraycr}
\makeatother
\newcolumntype{+}{>{\global\let\currentrowstyle\relax}}
\newcolumntype{^}{>{\currentrowstyle}}

\usepackage{setspace}

\newlength{\bracewidth}

\def\fudge{\mathchoice{}{}{\mkern.5mu}{\mkern.8mu}}
\def\bbc#1#2{{\rm \mkern#2mu\vbar\mkern-#2mu#1}}
\def\bbb#1{{\rm I\mkern-3.5mu #1}}
\def\bba#1#2{{\rm #1\mkern-#2mu\fudge #1}}
\def\bb#1{{\count4=`#1 \advance\count4by-64 \ifcase\count4\or\bba A{11.5}\or
   \bbb B\or\bbc C{5}\or\bbb D\or\bbb E\or\bbb F \or\bbc G{5}\or\bbb H\or
   \bbb I\or\bbc J{3}\or\bbb K\or\bbb L \or\bbb M\or\bbb N\or\bbc O{5} \or
   \bbb P\or\bbc Q{5}\or\bbb R\or\bbc S{4.2}\or\bba T{10.5}\or\bbc U{5}\or
   \bba V{12}\or\bba W{16.5}\or\bba X{11}\or\bba Y{11.7}\or\bba Z{7.5}\fi}}

\restylefloat{float}

\begin{document}


\title{Role of Mobility and Health Disparities on the Transmission Dynamics of Tuberculosis}

\author{Victor Moreno$^\dagger$, Baltazar Espinoza$^\dagger$, Kamal Barley$^\dagger$, Marlio Paredes$^\dagger$,\\  Derdei Bichara$^\dagger$, Anuj Mubayi$^\dagger$ and Carlos Castillo-Chavez$^\dagger$\\
\small{$^\dagger$ Simon A. Levin  Mathematical, Computational and Modeling Science Center,} \\
\small{Arizona State University, Tempe, AZ 85287}\\
\small{E-mails: \{vmmoren4, bespino6, kbarley, mparede4, dbichara, amubayi, ccchavez\}@asu.edu}
}

\date{}

\maketitle

\noindent
\textbf{Abstract:} The transmission dynamics of Tuberculosis (TB) involve complex epidemiological and socio-economical interactions between individuals living in highly distinct regional conditions. The level of exogenous reinfection and first time infection rates within high-incidence settings may influence the impact of control programs on TB prevalence. This study aims at enhancing the understanding of TB dynamics  via the study of scenarios, within {\it simplified}, two patch, risk-defined environments, in the presence of short term mobility and variations in reinfection and infection rates. The modeling framework captures the role of individuals' `daily' dynamics within and between places of residency, work or business via the  proportion of time spent in residence and as visitors to TB-risk environments (patches). As a result, the {\it effective population size} of Patch $i$ (home of $i$-residents) at time $t$ must account for visitors and residents of Patch $i$, at time $t$. The impact that {\it effective population size} and the distribution of {\it individuals' residence times} in different patches have on TB transmission and control are studied using selected scenarios where risk is defined by the estimated or perceive first time infection and/or exogenous re-infection rates. Our results suggest that, under certain conditions, allowing infected individuals to move from high to low TB prevalence areas (for example via the sharing of treatment and isolation facilities) may lead to a reduction in the total TB prevalence in the overall, here two-patch, population.




\section{Introduction}

Tuberculosis (TB), a communicable disease caused by bacteria (\textit{Mycobacterium tuberculosis}) remains among one of the leading causes of death worldwide. According to the World Health Organization's (WHO) report, 9.6 million people developed symptomatic TB infections resulting in 1.5 million TB-associated deaths in 2014 \cite{WHO2015}. Despite the existence of treatment and vaccine, it is estimated that one-third of the world population serves as TB reservoirs. The majority of these latently infected individuals live in developing countries where they are exposed to multiple TB risk factors. Individuals living in rural areas, mainly in developing countries, and in general below the poverty line disproportionately contribute to the documented TB burden \cite{Legesse2010, Ahn2005}. Data has shown strong association between poverty and TB, primarily in economically underprivileged countries \cite{Bhatt2009}. Vulnerable groups are at greater risk of TB infection compared with the general population because of overcrowding of individuals and substandard living or working conditions, poor nutrition, intercurrent diseases, and migration from (or to) higher-risk communities or nations are other known risk factors for TB \cite{ Ahn2005}. The Worldwide TB incidence rates seemed to have peaked (2004) after the HIV epidemic (1997) and then decreased at a rate of less than 1\% per year. Nonetheless, the overall worldwide TB-burden continues to rise as the world population continues to grow rapidly \cite{Lawn2011}. In addition, inappropriate treatment and the use of poor quality drugs have led to wild and antibiotic resistant strains contributing to the already high levels of TB-active incidence in recent years and  making TB a major global public health threat.

Gomes et \textit{al.} \cite{Gomes2012} found that TB-reinfection rates, that is, reinfection after successful treatment, are higher than TB infection rates among those with no prior TB-experience. In their model, they propose two mechanisms (for ongoing high prevalence in some regions): (i) past infections increase susceptibility to reinfection (ii) differences in susceptibility to infection contribute to increased re-infection rates among the treated. The study of these possibilities suggests that the last mechanism may be better supported by data. Gomes et \textit{al.} \cite{Gomes2012} noted that, hence, it is not surprising that rates of reinfection are higher at the population than at the individual level.

Metapopulation type TB transmission models offer a powerful set up for the study of the dynamics of TB infected individuals, on which the effectiveness of population-level TB interventions like treatment, movement restrictions, and local control measures can be studied. Prior TB-related studies have estimated TB incidence growth rates, explored the impact of interventions aimed at reducing TB prevalence and the impact of exogenous reinfection on TB dynamics, however, movement of individuals were ignored in such models.

Limited TB studies have considered models incorporating movement via mass transportation \cite{Castillo-Chavez2000}, or taking into account the impact of sudden blips of immigration, which may be central to TB re-emergence \cite{Tewa2012, Liu2012785, Zhou2008, BraVdd01, shim2006note}, or that account for co- infections, specially with HIV \cite{Kapitanov2015, Nthiiri2015, Bhunu2009, Bowong2010, Hohmann2013, Roeger2009a}, or that account for relapse \cite{Gomes2012, Millet2013, Marx2014, Luzze2013, Tiemersma:2011bk}, or that account for antibiotic, drug, and ultra-drug resistance \cite{Okuonghae2013, Ozcaglar2012, Bhunu2011, Lipsitch1998, Agusto2014, Cohen2009}, or models that account for TB re-activation and progression \cite{Feng2000, Cohen2007, zheng2014modeling}. In addition, models assuming negligible immigration might not capture the real dynamics of tuberculosis in open populations when high levels of diversity is caused by immigrants \cite{Ozcaglar2012}.

Research aimed at increasing the understanding of the transmission dynamics of TB that explicitly incorporate the role of heterogeneous TB-risk environments is limited. The goal of this study is to understand the impact of residence times and population sizes, across distinct risk environments, on the TB transmission dynamics when risk being defined in terms of new infection and/or exogenous infection rates. We define residence time in a place as the average proportion of daily time an individual spends in a given region or patch. In particular, we address three questions 
(i) \textit{How does mobility changes TB prevalence via the tradeoff between exogenous and direct first time infection rates?}, 
(ii) \textit{How differences in population sizes of the patches can influence the impact of mobility on total infections?} and 
(iii) \textit{Which among the two, first time infection rates and exogenous reinfection rates, is capable of sustaining higher TB prevalence?}

\section{Method: TB Dynamic Model}\label{sec:ModelDerivation}

We consider a  model for the  transmission dynamics of TB in populations interacting in two distinct regions/patches. First, we introduce a model with one patch and then extend it to capture two patches by explicitly incorporating short term movement of individuals between and within patches. The two-patch mobility model is used to address the role of movement and patch-risk on TB dynamics.

\begin{table}[ht]
\scriptsize
\centering
\label{tab:defscen}
\begin{tabular}{|l|l|}
\hline
\multicolumn{2}{|c|}{\cellcolor[HTML]{C0C0C0}\textbf{Nomenclature}}                                                                                                                                                                                                                                                                             \\ \hline
\textbf{Risk}                                                                                                      & \begin{tabular}[c]{@{}l@{}}Interpreted based on levels of infection rate, prevalence, \\  
or average contacts (via population size)\end{tabular}                                                                            \\ \hline
\textbf{High-risk patch}                                                                                           & \begin{tabular}[c]{@{}l@{}}Defined either by high direct first time infection rate (i.e., high $\beta$ \\ 
 or high corresponding $\mathcal{R}_0$) or by high exogenous infection rate \\ (i.e., high $\delta$)\end{tabular} \\ 
\hline
\textbf{\begin{tabular}[c]{@{}l@{}}Enhanced socio-economic conditions \\ (reducing health disparity)\end{tabular}} & \begin{tabular}[c]{@{}l@{}}Defined by better healthcare infrastructure which is incorporated 
\\  by high prevalence of a disease (i.e., high $I(0)$)  in a large \\ 
population (i.e., large$N$)\end{tabular}                 \\ \hline
\textbf{Mobility}                                                                                                  & \begin{tabular}[c]{@{}l@{}}Captured by average residence times of an individual \\ in different patches (i.e.,  by using $\mathbb{P}$ matrix)\end{tabular}                                                                 \\ \hline
\multicolumn{2}{|c|}{\cellcolor[HTML]{C0C0C0}\textbf{\begin{tabular}[c]{@{}c@{}}Scenarios (assume high-risk and enhanced socio-economic conditions in \\ Patch 1 as compared to Patch 2)\end{tabular}}}                                                                                                                                         \\ \hline
\textbf{Scenario 1}                                                                                                & $\underbrace{\beta_1 > \beta_2, \ \delta_1 = \delta_2}_{\text{high risk}}$; \quad $\underbrace{\dfrac{I_1(0)}{N_1} > \dfrac{I_2(0)}{N_2}, \ N_1 > N_2;}_{\text{enhanced socio-economic conditions}}$ \quad $\underbrace{\text{vary} \ p_{12}}_{\text{mobility}}$                                                                                                        \\ \hline
\textbf{Scenario 2}                                                                                                & $\underbrace{\beta_1 = \beta_2, \ \delta_1 > \delta_2}_{\text{high risk}}$; \quad $\underbrace{\dfrac{I_1(0)}{N_1} > \dfrac{I_2(0)}{N_2}, \ N_1 > N_2;}_{\text{enhanced socio-economic conditions}}$ \quad $\underbrace{\text{vary} \ p_{12}}_{\text{mobility}}$                                                                                                                                                                                                               \\ \hline
\end{tabular}
\caption{Definitions and scenarios in the study}
\end{table}

\subsection{Simple TB model for one patch}\label{sec:twopatch}

The transmission dynamics of TB in homogeneously mixing populations is represented by systems of differential equations describing TB contagion. The population in the model is divided into three sub-populations each corresponding to an epidemiological TB state: susceptible individuals ($S$), noninfectious infected, that is, latent individuals ($L$), and actively infectious individuals ($I$).

The model considers two contagion pathways: direct progression (fast dynamics) and endogenous reactivation (slow progression, often years after infection). Susceptible individuals ($S$) may get infected through contacts with individuals with active-infections ($I$), moving to either the noninfectious latent class ($L$) or the actively infectious ($I$) state. The fraction $(1-q)$ denotes the proportion of infected individuals that move directly into the infectious stage ($I$). Reactivation from longstanding latent infections is modeled by the transition of individuals from the noninfectious to the infectious state (progression to active TB) via  endogenous reactivation (at the per capita rate $\gamma$), or via exogenous reinfection.  Infectious individuals may be  treated at the per capita rate $\rho$ moving into the non-infectious infected category $L$ as total {\it mycobacterium} elimination is assumed to be non possible.

\begin{figure}[H]
 \centering
 \includegraphics[width=0.5\textwidth]{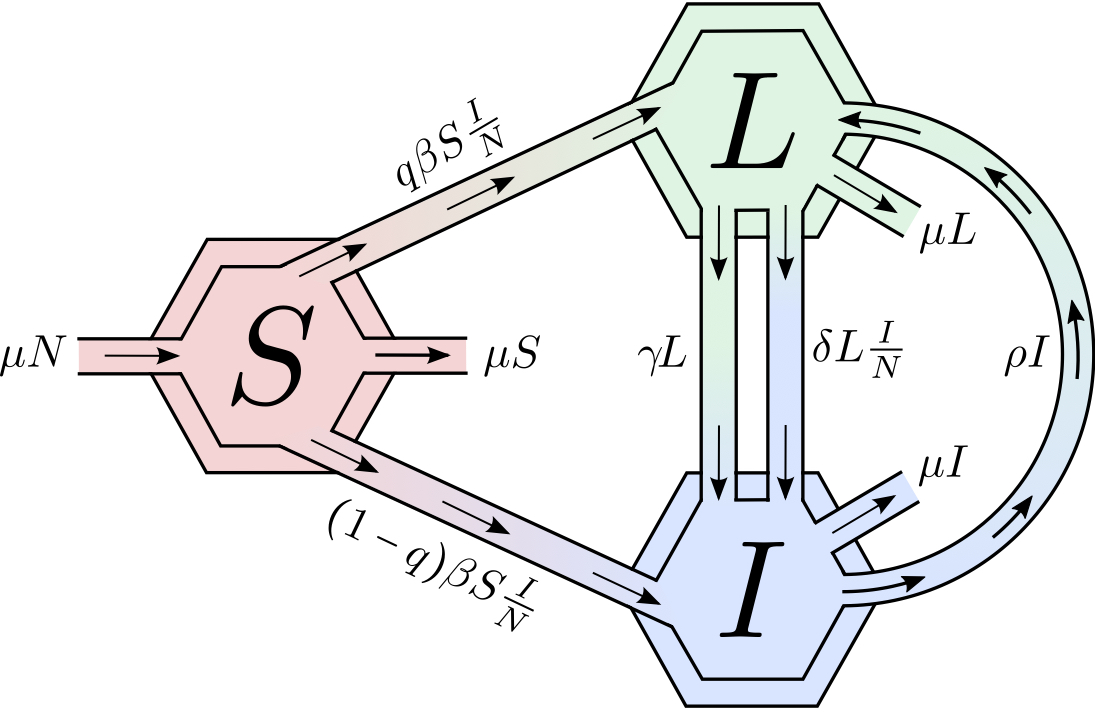} 
 \caption{Flow diagram between the three compartments of the model: susceptible ($S$), infected latent ($L$), infectious ($I$)} 
 \label{flow_tb}
\end{figure}

The model assumes that (1) the population  is constant; (2) TB-induced deaths are negligible and hence ignored; (3) a fraction of individuals are infectious; (4) individuals may control an active infection without treatment moving back to the latent class; (5) individuals in the latent class may relapse and develop active TB  or remain in the class until  death due to natural causes (that is, not TB). Figure \ref{flow_tb} shows the flow diagram associated with the transmission dynamics of the TB model used .

This model follows the structure in \cite{Feng2000,Mccluskey2006,Zheng2014} where exogenous reinfection, fast and slow progression are considered. The basic reproduction number and the existence of a parameters' range for which there are two stable equilibria, disease free and endemic steady states are highlighted in  \cite{Feng2000,Mccluskey2006,Zheng2014}. The basic reproduction number of the model is given by
\[ \mathcal R_0=\frac{\beta (\gamma+(1-q)\mu)}{\mu(\mu+\rho+\gamma)}\label{R01Patch}\tag{$\star$}\]
The basic reproduction number ($\mathcal R_0 $) gives the average number of secondary infections generated by a typically infected individual in a population of susceptible individuals. In the presence of exogenous reinfection, excluding  fast progression ($q=1$ and $\delta>0$), it is known that the model can support  two stable equilibria \cite{Feng2000}. The role of TB, in this case would be closely linked not only to $\mathcal{R}_0$ but also to the initial conditions.  We proceed to build a two-patch model, under a residency-time matrix, using the model outlined above.

\subsection{Heterogeneity through residence times}
\label{sec:twopatch}

Let $N_1$ and $N_2$ be the host population of Patch 1 and 2, respectively. The population of Patch 1 spends, on the average, the proportion $p_{11}$  of its time in residency, that is, in Patch 1 and so, the proportion $p_{12}$  of its time in Patch 2 ($p_{11}+p_{12}=1$).  Similarly,  residents of Patch 2 spend the proportion $p_{22}$  of time  their time  in Patch  2 and  $p_{21}=1-p_{22}$ in Patch 1. Hence, at time $t$, the {\textit effective population} in Patch 1 is $p_{11}N_1+p_{21}N_2$ while the {\textit effective population} of Patch 2, at time $t$, is $p_{12}N_{1}+p_{22}N_{2}$. The susceptible population of Patch 1 ($S_1$) may become infected in Patch 1 (if current in Patch 1, i.e. $p_{11}S_1$) or in Patch 2 (if current in Patch 2, i.e. $p_{12}S_2$). In short, from this Lagrangian approach to capture movement of individuals, we conclude that the {\textit effective proportion of infectious} individuals in Patch 1 at time $t$ is 

$$\frac{p_{11}I_{1}+p_{21}I_{2}}{p_{11}N_{1}+p_{21}N_{2}}.$$

\begin{figure}[h]
\centering
\includegraphics[scale =.15]{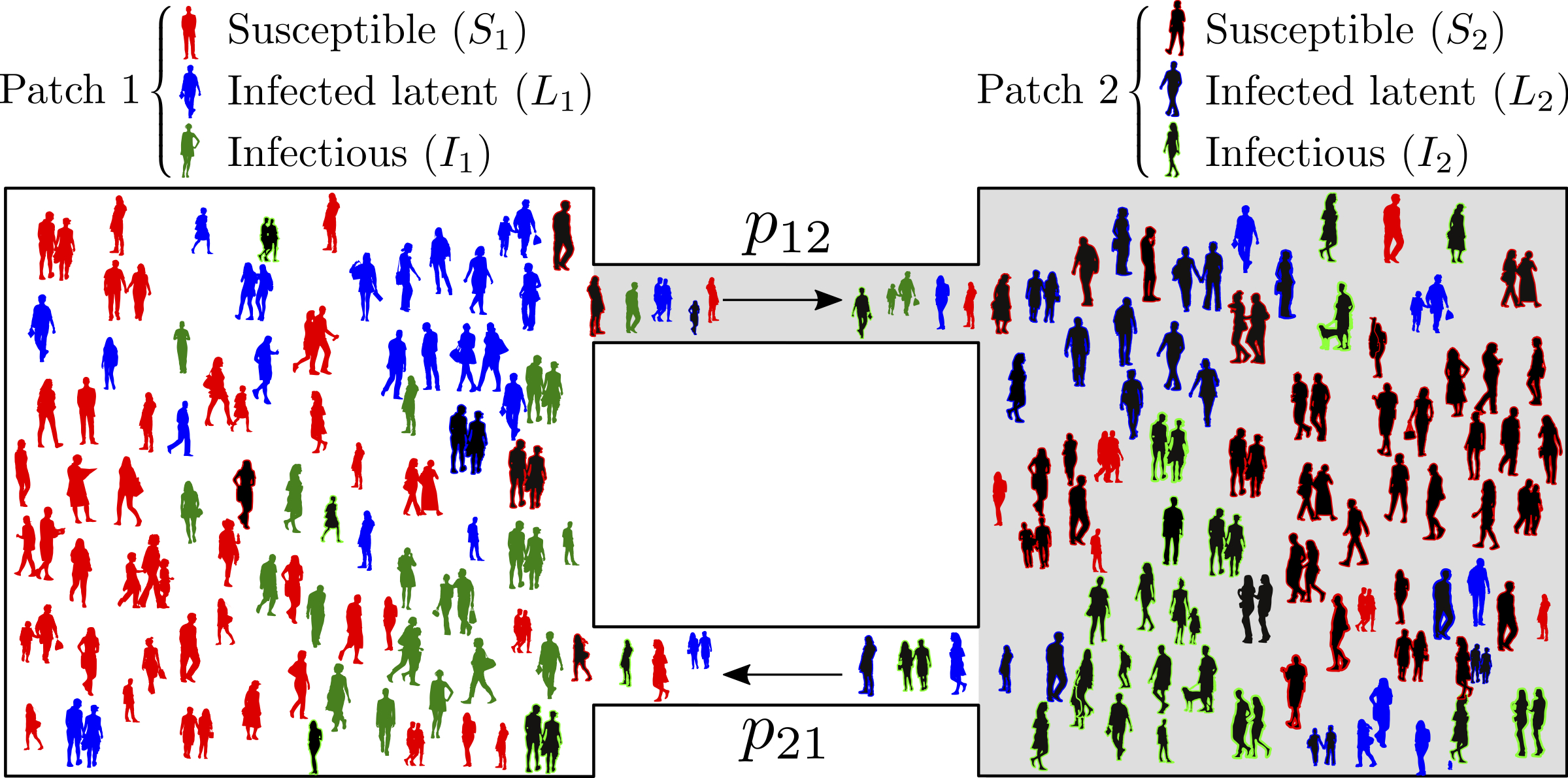}
\caption{Schematic description of the Lagrangian approach between two patches.}
\label{fig:lag}
\end{figure}

Thus, the dynamics of infection among susceptible, resident  individuals of Patch 1  is given by 
\begin{equation}\label{EqS1}
\dot S_{1}=\mu_{1}N_{1}-\beta_1 p_{11}S_{1}\frac{p_{11}I_{1}+p_{21}I_{2}}{p_{11}N_{1}+p_{21}N_{2}}-\beta_{2}p_{12}S_1\frac{p_{12}I_{1}+p_{22}I_{2}}{p_{12}N_{1}+p_{22}N_{2}}-\mu_1 S_{1}.\end{equation} 

The dynamics of the Patch 1 residents acquiring latent, asymptomatic infections, is, 
\begin{multline}\label{EqE1}
\dot{L}_{1}=q\beta_1 p_{11}S_{1}\frac{p_{11}I_{1}+p_{21}I_{2}}{p_{11}N_{1}+p_{21}N_{2}}+q\beta_{2}p_{12}S_1\frac{p_{12}I_{1}+p_{22}I_{2}}{p_{12}N_{1}+p_{22}N_{2}}\\
-\delta_1 p_{11}L_{1}\frac{p_{11}I_{1}+p_{21}I_{h,2}}{p_{11}N_{1}+p_{21}N_{2}}-\delta_{2}p_{12}L_1\frac{p_{12}I_{1}+p_{22}I_{2}}{p_{12}N_{1}+p_{22}N_{2}}-(\mu_1+\gamma_1) I_{1}-\rho_1 I_1, \end{multline}
and the dynamics of the Patch 1 residents becoming infectious is 
\begin{multline}\label{EqI1}
\dot{I}_{1}=(1-q)\beta_1 p_{11}S_{1}\frac{p_{11}I_{1}+p_{21}I_{2}}{p_{11}N_{1}+p_{21}N_{2}}+(1-q)\beta_{2}p_{12}S_1\frac{p_{12}I_{1}+p_{22}I_{2}}{p_{12}N_{1}+p_{22}N_{2}}\\
+\delta_1 p_{11}L_{1}\frac{p_{11}I_{1}+p_{21}I_{h,2}}{p_{11}N_{1}+p_{21}N_{2}}+\delta_{2}p_{12}L_1\frac{p_{12}I_{1}+p_{22}I_{2}}{p_{12}N_{1}+p_{22}N_{2}}-(\mu_1+\gamma_1) I_{1}+\rho_1 I_1.\end{multline}

The use of (\ref{EqS1}), (\ref{EqE1}),(\ref{EqI1}) determines the complete dynamics of TB, in two patches, and it is given by the following System( $i=1,2$):
\begin{equation} \label{PatchGenConsR}
\left\{\begin{array}{llll}
\dot S_{i}=\mu_{i}N_{i}-\sum_{j=1}^2\beta_j p_{ij}S_{i}\frac{ \sum_{k=1}^{2}p_{kj}I_{k}}{ \sum_{k=1}^{2}p_{kj}N_{k}}-\mu_i S_{i},\\
\dot{L}_{i}=q\sum_{j=1}^2\beta_j p_{ij}S_{i}\frac{ \sum_{k=1}^{2}p_{kj}I_{k}}{ \sum_{k=1}^{2}p_{kj}N_{k}}-\sum_{j=1}^2\delta_j p_{ij}L_{i}\frac{ \sum_{k=1}^{2}p_{kj}I_{k}}{ \sum_{k=1}^{2}p_{kj}N_{k}}-(\gamma_i+\mu_i)L_{i}+\rho_i I_i,\\
\dot{I}_{i}=(1-q)\sum_{j=1}^2\beta_j p_{ij}S_{i}\frac{ \sum_{k=1}^{2}p_{kj}I_{k}}{ \sum_{k=1}^{2}p_{kj}N_{k}}+\sum_{j=1}^2\delta_j p_{ij}L_{i}\frac{ \sum_{k=1}^{2}p_{kj}I_{k}}{ \sum_{k=1}^{2}p_{kj}N_{k}}+\gamma L_{i}-(\mu_i+\rho_i) I_{i}.
\end{array}\right.
\end{equation}

Let $N_i=S_i+L_i+I_i$ the total population of Patch $i$, $i=1,2$. Since the population in each Patch is constant, System (\ref{PatchGenConsR}) has the same qualitative dynamics than the following reduced system since the  total host population  is constant:
\begin{equation} \label{PatchGenFinal}
\left\{\begin{array}{llll}
\dot{L}_{i}=q\sum_{j=1}^2\beta_j p_{ij}(N_i-L_{i}-I_i)\frac{ \sum_{k=1}^{2}p_{kj}I_{k}}{ \sum_{k=1}^{2}p_{kj}N_{k}}-\sum_{j=1}^2\delta_j p_{ij}L_{i}\frac{ \sum_{k=1}^{2}p_{kj}I_{k}}{ \sum_{k=1}^{2}p_{kj}N_{k}}-(\gamma_i+\mu_i)L_{i}+\rho_i I_i,\\
\dot{I}_{i}=\sum_{j=1}^2p_{ij}\left((1-q)\beta_j (N_i-L_{i}-I_i)+\delta_j L_{i}\frac{}{}\right)\frac{ \sum_{k=1}^{2}p_{kj}I_{k}}{ \sum_{k=1}^{2}p_{kj}N_{k}}+\gamma L_{i}-(\mu_i+\rho_i) I_{i}.
\end{array}\right.
\end{equation}
A schematic description of the two-patch dynamical model is provided in Figure \ref{fig:lag}.

\begin{table}[H]
\centering
\label{tab:Param} 
\resizebox{\textwidth}{!}{\rowcolors{2}{gray!10}{white}
\begin{tabular}{c>{\raggedright}p{4.1in}llr}
\hline 
\rowcolor{gray!30} \textbf{Parameters }  & \textbf{Description }  && Ranges(units)  & References\tabularnewline
\hline 
$\beta_{i}$  & Susceptibility to TB invasion in Patch $i$  && 0.01 - 0.0192 ($y^{-1}$)  & \cite{Cohen2007} \tabularnewline
$\delta_{i}$  & Susceptibility to exogenous TB progression in Patch $i$  && 0.0026 - 0.0053 ($y^{-1}$)  & \cite{Blower1995} \tabularnewline
$\mu_{i}$  & Natural birth and death (per capita)  && 0.0104 - 0.0143 ($y^{-1}$) & \cite{Gomes2004} \tabularnewline
$\rho$  & Relapse (per capita)  && 0.0010 - 0.0083($y^{-1}$)  & \cite{Langley2014, Dowdy2012,Langley2014} \tabularnewline
$\gamma_{i}$  & Activation from latency in Patch $i$ (per capita) && 0.0017 - 0.0036 ($y^{-1}$) & \cite{Okuonghae2013} \tabularnewline
$q$  & Proportion of individuals that develop latent TB  && 0.9 (dimensionless)  & \cite{Gomes2004} \tabularnewline
$p_{ij}$  & Proportion of time that residents of Patch $i$ spend in Patch $j$  && Varies (dimensionless)  & -- \tabularnewline
\bottomrule
\end{tabular}} 
\par
\centering{}\protect\protect\caption{Description of the parameters used in System (\ref{PatchGenFinal}).}
\end{table}

\section{Results}

\subsection{Model Analysis}
The disease-free equilibrium of (\ref{PatchGenFinal}) is located at the origin of the positive orthan $\mathbb{R}^4_+$, that is $E_0=0_{\mathbb{R}^4_+}$. The basic reproduction number of Model (\ref{PatchGenFinal}) is $\mathcal R_0$ is computed following the next generation method \cite{van2002reproduction, diekmann1990definition}. We decompose System  (\ref{PatchGenFinal}) into a sum of ``new infection'' vector, denoted by $\mathcal F$, and ``transition'' vector, denoted by $\mathcal V$. Hence,

$$
\begin{array}{rcl}
\begin{bmatrix}
\dot L_1 \\
\dot L_2\\
\dot E_1  \\
\dot E_2
\end{bmatrix} & = & \mathcal F+\mathcal V\nonumber \\
    & = &
\begin{bmatrix}
q\sum_{j=1}^2\beta_j p_{1j}(N_1-L_{1}-I_1)\frac{ \sum_{k=1}^{2}p_{kj}I_{k}}{ \sum_{k=1}^{2}p_{kj}N_{k}}\\
q\sum_{j=1}^2\beta_j p_{2j}(N_2-L_{2}-I_2)\frac{ \sum_{k=1}^{2}p_{kj}I_{k}}{ \sum_{k=1}^{2}p_{kj}N_{k}}\\
(1-q)\sum_{j=1}^2\beta_j p_{1j}(N_1-L_{1}-I_1)\frac{ \sum_{k=1}^{2}p_{kj}I_{k}}{ \sum_{k=1}^{2}p_{kj}N_{k}}\\
(1-q)\sum_{j=1}^2\beta_j p_{2j}(N_2-L_{2}-I_2)\frac{ \sum_{k=1}^{2}p_{kj}I_{k}}{ \sum_{k=1}^{2}p_{kj}N_{k}}
 \end{bmatrix} + \\ [40pt]
  &  & + 
\begin{bmatrix}
-\sum_{j=1}^2\delta_j p_{1j}L_{1}\frac{ \sum_{k=1}^{2}p_{kj}I_{k}}{ \sum_{k=1}^{2}p_{kj}N_{k}}-(\gamma_1+\mu_1)L_{1}+\rho_1 I_1\\
-\sum_{j=1}^2\delta_j p_{2j}L_{2}\frac{ \sum_{k=1}^{2}p_{kj}I_{k}}{ \sum_{k=1}^{2}p_{kj}N_{k}}-(\gamma_2+\mu_2)L_{2}+\rho_2 I_2\\
\sum_{j=1}^2p_{1j}\delta_j L_{1}\frac{ \sum_{k=1}^{2}p_{kj}I_{k}}{ \sum_{k=1}^{2}p_{kj}N_{k}}+\gamma L_{1}-(\mu_1+\rho_1) I_{1}\\
\sum_{j=1}^2p_{2j}\delta_j L_{2}\frac{ \sum_{k=1}^{2}p_{kj}I_{k}}{ \sum_{k=1}^{2}p_{kj}N_{k}}+\gamma L_{2}-(\mu_2+\rho_2) I_{2}
\end{bmatrix}
\end{array}
$$

The rationale behind the presence of nonlinear terms, which represent the infectiousness of latent by infectious individuals, in the $\mathcal V$ vector is that these terms do not, technically, represent ``new infection''. By denoting $F$ and $V$, the Jacobian matrices of $\mathcal F$ and $\mathcal V$ respectively, evaluated at the disease free equilibrium $E_0$, the basic reproduction number is the spectral radius of the next generation matrix $-FV^{-1}$ \cite{van2002reproduction, diekmann1990definition}. Hence,
$\mathcal R_0=\rho(-FV^{-1})$ where
$$
-FV^{-1}=\begin{bmatrix}
q\gamma_1k_{11} & q\gamma_2k_{12} & q(\mu_1+\gamma_1)k_{11} & q(\mu_2+\gamma_2)k_{21} \\
q\gamma_1k_{21}& q\gamma_2k_{22} & q(\mu_1+\gamma_1)k_{21} & q(\mu_2+\gamma_2)k_{22}\\
(1-q)\gamma_1k_{11} & (1-q)\gamma_2k_{12} & (1-q)(\mu_1+\gamma_1)k_{11} & (1-q)(\mu_2+\gamma_2)k_{12}  \\
(1-q)\gamma_1k_{21}& (1-q)\gamma_2k_{22} & (1-q)(\mu_1+\gamma_1)k_{21} & (1-q)(\mu_2+\gamma_2)k_{22}
 \end{bmatrix}$$
where
\begin{align*}
k_{11}=&\left(\frac{\beta_1p^2_{11}N_1}{p_{11}N_1+p_{21}N_2}+\frac{\beta_2p^2_{12}N_1}{p_{12}N_1+p_{22}N_2}\right)\frac{1}{\mu_1(\gamma_1+\mu_1+\rho_1)}\\
=&\left(\frac{\beta_1p^2_{11}N_1}{p_{11}N_1+p_{21}N_2}+\frac{\beta_2p^2_{12}N_1}{p_{12}N_1+p_{22}N_2}\right)\frac{\mathcal{R}_{0}^{1}}{\beta_1(\gamma_1+(1-q)\mu_1)},
\end{align*}

\begin{align*}
k_{12}=&\left(\frac{\beta_1p_{11}p_{21}N_1}{p_{11}N_1+p_{21}N_2}+\frac{\beta_2p_{12}p_{22}N_1}{p_{12}N_1+p_{22}N_2}\right)\frac{1}{\mu_2(\gamma_2+\mu_2+\rho_2)}\\
=&\left(\frac{\beta_1p_{11}p_{21}N_1}{p_{11}N_1+p_{21}N_2}+\frac{\beta_2p_{12}p_{22}N_1}{p_{12}N_1+p_{22}N_2}\right)\frac{\mathcal{R}_{0}^{2}}{\beta_2(\gamma_2+(1-q)\mu_2)},
\end{align*}

\begin{align*}
k_{21}=&\left(\frac{\beta_1p_{11}p_{21}N_2}{p_{11}N_1+p_{21}N_2}+\frac{\beta_2p_{12}p_{22}N_2}{p_{12}N_1+p_{22}N_2}\right)\frac{1}{\mu_1(\gamma_1+\mu_1+\rho_1)}\\
=&\left(\frac{\beta_1p_{11}p_{21}N_2}{p_{11}N_1+p_{21}N_2}+\frac{\beta_2p_{12}p_{22}N_2}{p_{12}N_1+p_{22}N_2}\right)\frac{\mathcal{R}_{0}^{1}}{\beta_1(\gamma_1+(1-q)\mu_1)},
\end{align*}
and 
\begin{align*}
k_{22}=&\left(\frac{\beta_1p_{21}^2N_2}{p_{11}N_1+p_{21}N_2}+\frac{\beta_2p_{22}^2N_2}{p_{12}N_1+p_{22}N_2}\right)\frac{1}{\mu_2(\gamma_2+\mu_2+\rho_2)}\\
=&\left(\frac{\beta_1p_{21}^2N_2}{p_{11}N_1+p_{21}N_2}+\frac{\beta_2p_{22}^2N_2}{p_{12}N_1+p_{22}N_2}\right)\frac{\mathcal{R}_{0}^{2}}{\beta_2(\gamma_2+(1-q)\mu_2)}.
\end{align*}

Note that $\mathcal{R}_0=f(\mathbb{P},\mathcal{R}_{0}^1,\mathcal{R}_{0}^2)$ where $\mathcal{R}_{0}^1$ and $\mathcal{R}_{0}^2$ are the basic reproductive numbers of patch 1 and 2, respectively, when $p_{11}=1=p_{22}$, that is, when there is no movement. $\mathbb{P} = (p_{ij})_{1\leq i,j\leq2}$ is referred as the residence times matrix of the model. The corresponding expressions of $\mathcal R_0^1$ and $\mathcal R_0^2$ are given by (\ref{R01Patch}).

The analysis of the Model (\ref{PatchGenFinal}) suggests that the  disease dies out from both patches if $\mathcal R_0\leq1$ or persists in both patches otherwise for the case when $q=1$ and $\delta=0$ (i.e., in the absence of fast progression and exogenous infections because the residence times matrix becomes irreducible) (See \cite{bichara2015sis,BicharaCCC2015} for the mathematical proofs). We assume $q=1$ through out this study and $\delta>0$, numerical simulations suggest complex dynamics (i.e., multiple non-trivial equilibria) of the system.

\begin{figure}[H]
\centering
  \includegraphics[width=0.98\textwidth]{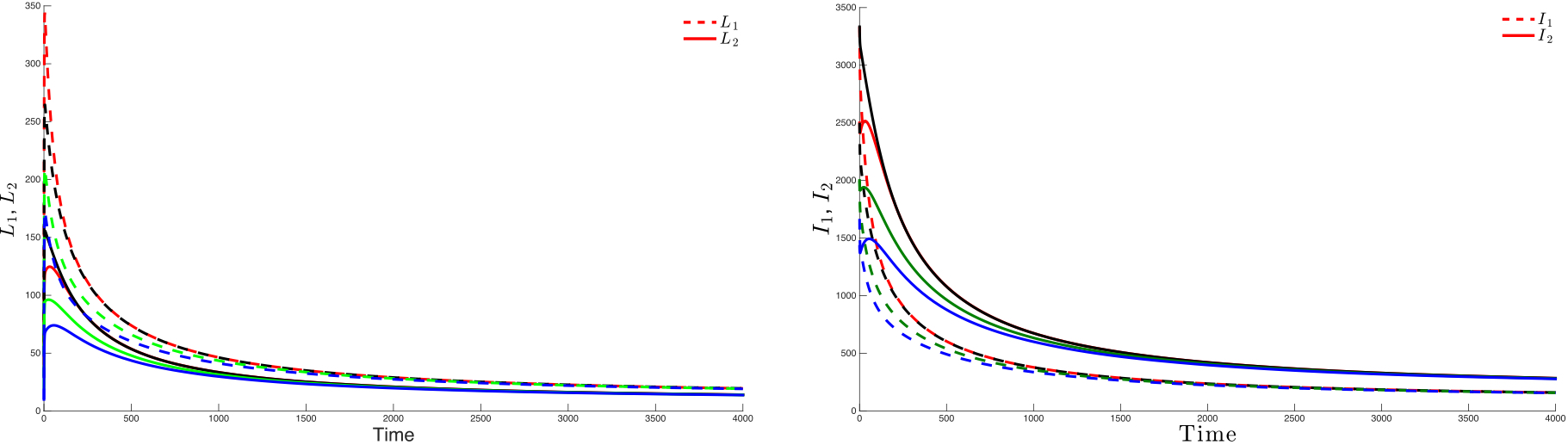}
\caption{Dynamics of infectious and latent if the two patches are strongly connected and  $\mathcal R_0>1$. For four different initial conditions, the latent (top) and infected (bottom) populations of Patch 1 and Patch 2 attain an endemic level if $\mathcal R_0>1$} \label{fig:twofigsRobustEE}
\end{figure}

\begin{figure}[H]
\centering
   \includegraphics[width=0.98\textwidth]{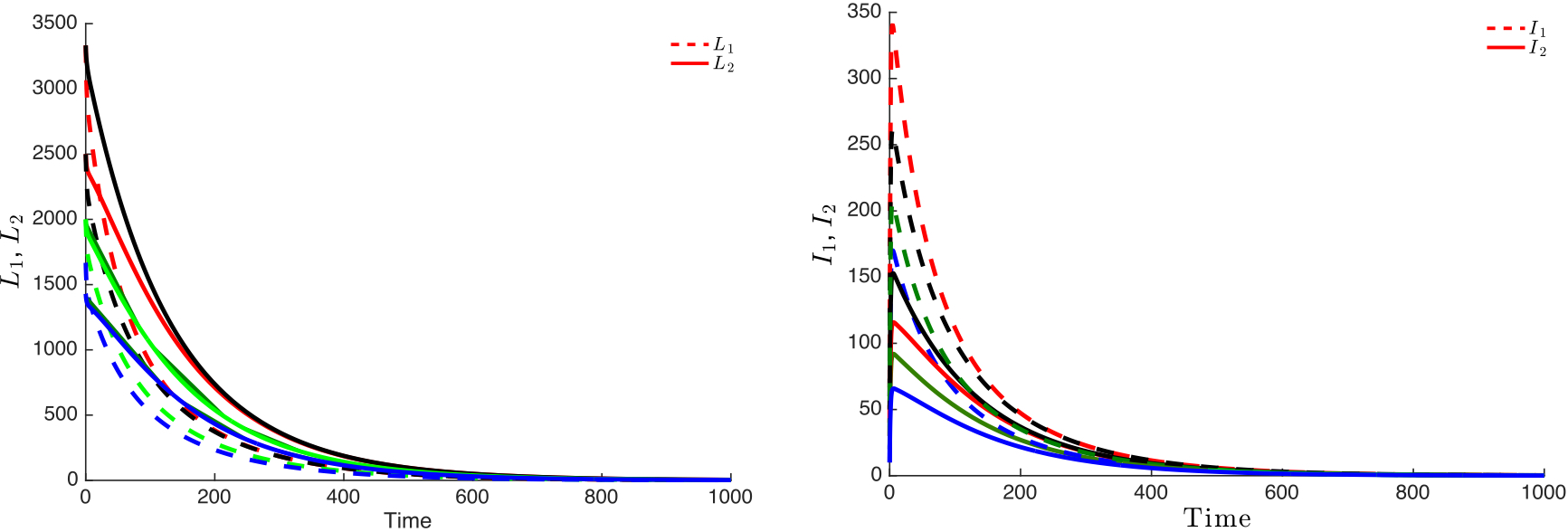}
\caption{The infectious and latent populations in the two patches converge to zero for four different initial conditions if $\mathcal R_0\leq1$.} \label{fig:twofigsRobustDFE}
\end{figure}

Figure \ref{fig:twofigsRobustEE} highlights this robustness, that is for four different initial conditions, the trajectories of the latently infected individuals  (Figure \ref{fig:twofigsRobustEE} left) as well as the  actively-infected (Figure \ref{fig:twofigsRobustEE} right) converge towards the endemic state as time becomes large. The case when $\mathcal R_0\leq1$, leads to the elimination of the disease from both  patches irrespective of the initial conditions as shown in Figure \ref{fig:twofigsRobustDFE}.

\begin{figure}[H]
\centering
   \includegraphics[width=0.95\textwidth]{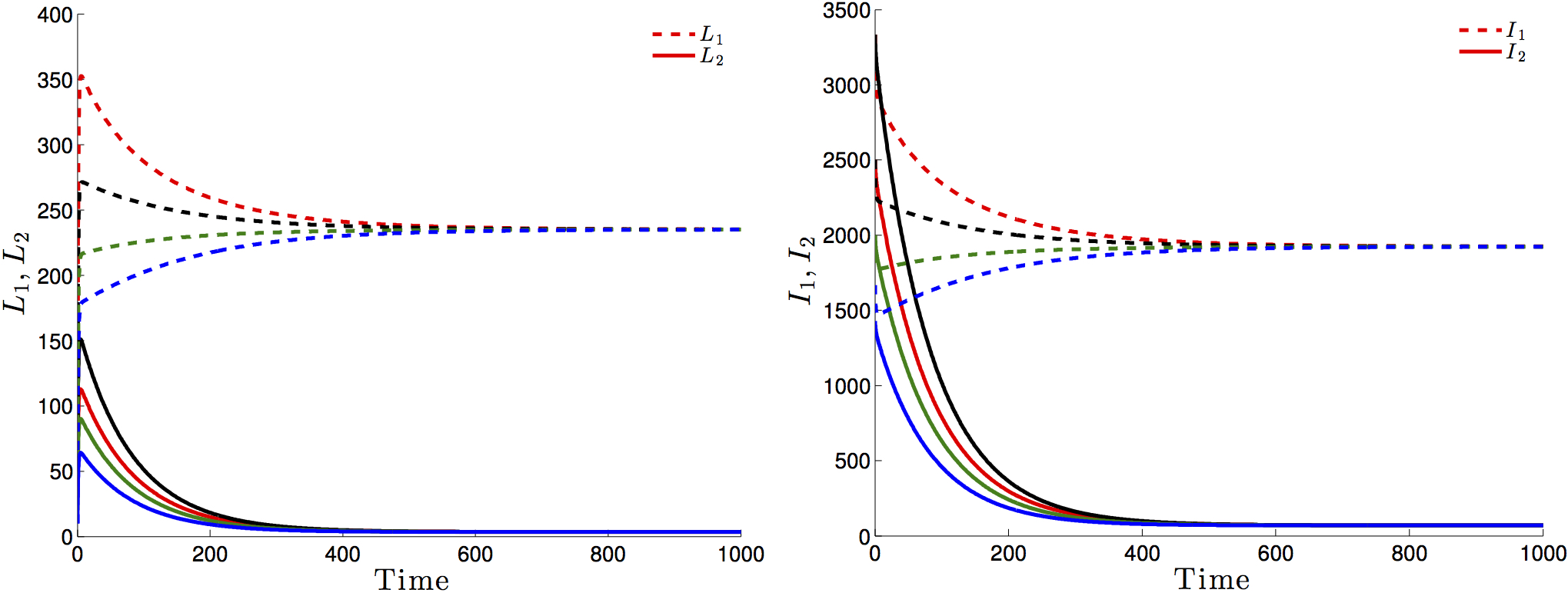}
\caption{Dynamics when the two patches are weakly connected and  $\mathcal R_0>1$. The latent (top) and infected (bottom) of both patches reach an endemic level but Patch 2 approaches a lower level of endemicity ($\mathcal R_0^1=1.4150$ and $\mathcal R_0^2=0.1417$ if completely isolated)} \label{EIAlmostIsolatedFig5}
\end{figure}

If Patch 1 is high risk (that is, $\mathcal{R}_0>1$) and, if the connectivity between the two patches is not strong ($p_{21}\approx 0$ and $p_{12}\approx0$), then the disease will persist in both patches, even though that the number of latently-infected and actively-infectious individuals in Patch 2 is small (See Figure \ref{EIAlmostIsolatedFig5} left and \ref{EIAlmostIsolatedFig5} right).

\vspace{1cm}

\begin{figure}[H]
\centering
   \includegraphics[width=0.95\textwidth]{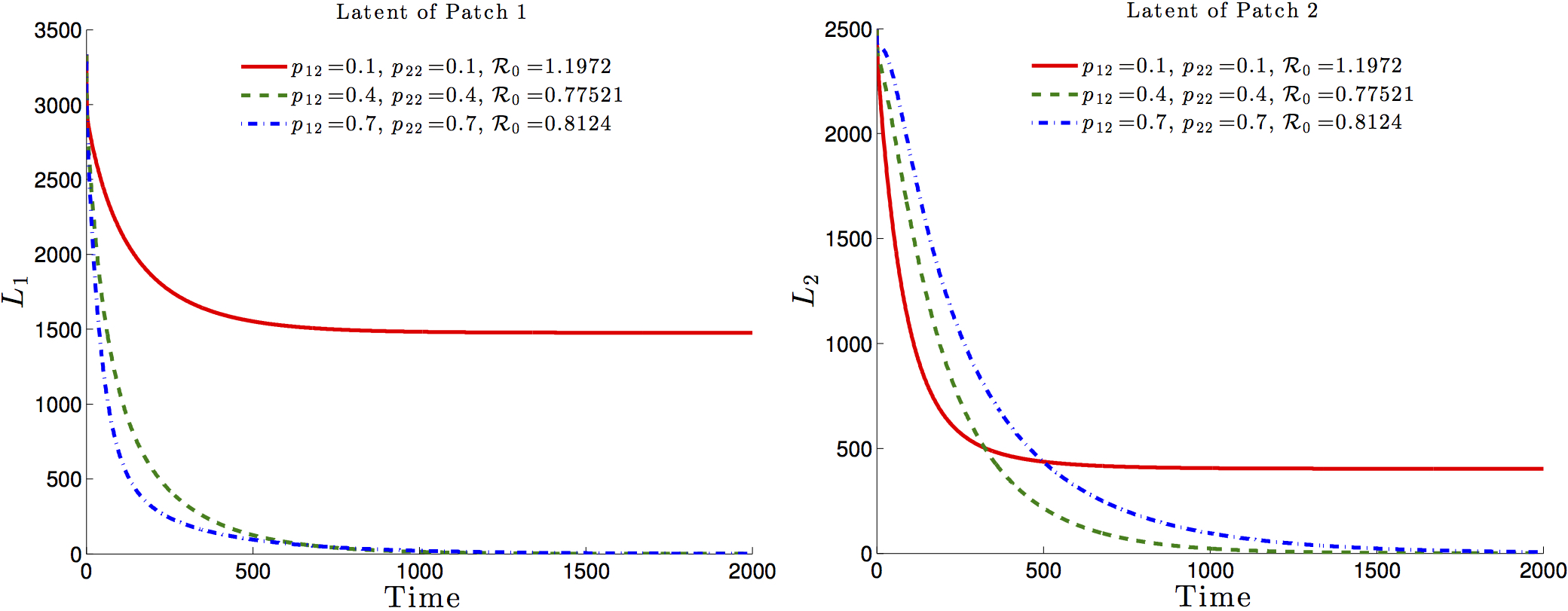}
\caption{Effects of the residence time matrix on the basic reproduction number and the disease dynamics. In both patches, the \textit{latent} TB populations go to zero if $\mathcal R_0(\mathbb P)<1$ and reach an endemic level if $\mathcal R_0(\mathbb P)>1$.} \label{fig:twofigsEffectsPonR0E1E2}
\end{figure}

\begin{figure}[H]
\centering
   \includegraphics[width=0.95\textwidth]{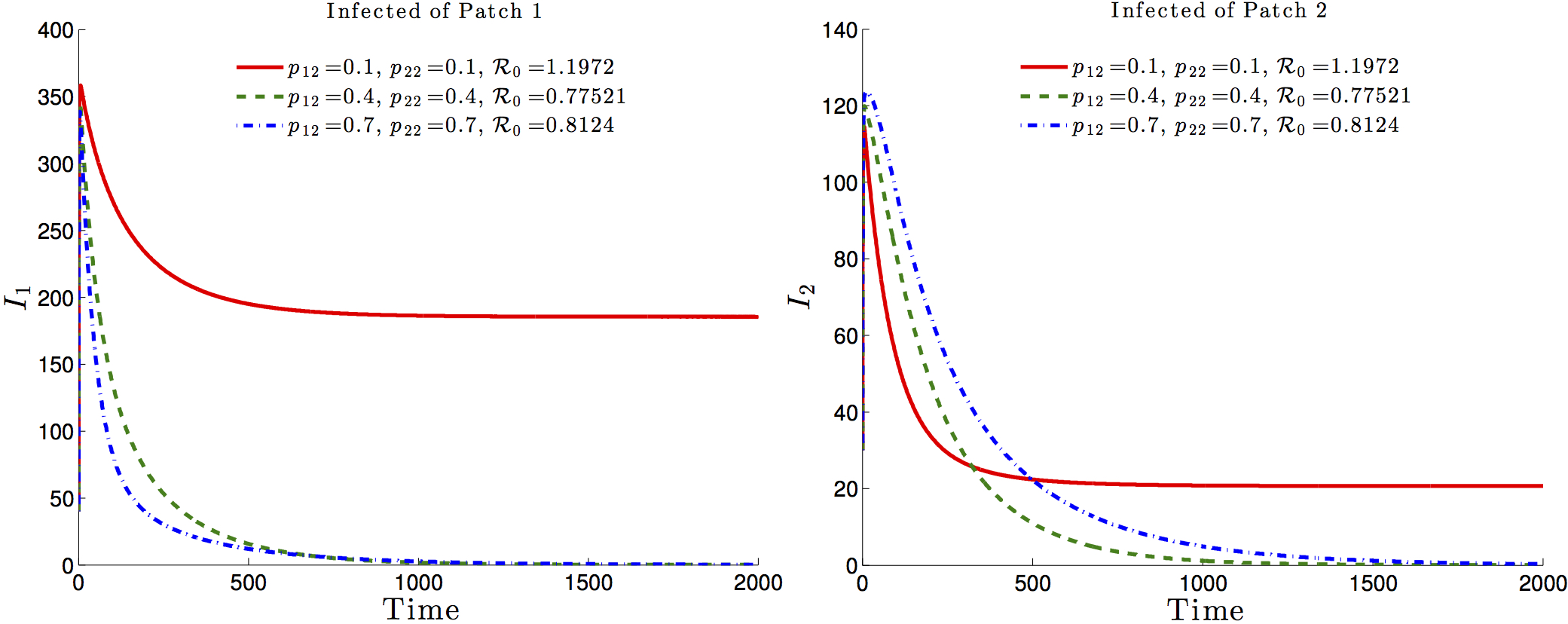}
\caption{Effects of the residence time matrix on the basic reproduction number and the disease dynamics. In both patches, the \textit{infected} TB populations go to zero if $\mathcal R_0(\mathbb P)<1$ and reach an endemic level if $\mathcal R_0(\mathbb P)>1$.}
 \label{fig:twofigsEffectsPonR0I1I2}
\end{figure}

The effects of the residence times matrix $\mathbb P = (p_{ij})_{1\leq i,j\leq2}$ on the basic reproduction number $\mathcal R_0(\mathbb P)$ and, consequently on the disease dynamics, are highlighted in Figure \ref{fig:twofigsEffectsPonR0E1E2}  and  Figure \ref{fig:twofigsEffectsPonR0I1I2}. It is observed that the basic reproduction number is a decreasing function of  $p_{12}$, i.e. the residence time of high risk residents (Patch 1 residents) in the low risk Patch 2. Such a decrease would ultimately drive the basic reproduction number to a value less than one with the latent and infected populations, under such mobility schedules going to zero in both patches (See Figure \ref{fig:twofigsEffectsPonR0E1E2} and Figure \ref{fig:twofigsEffectsPonR0I1I2}, dash-dotted green and dashed blue).

Now, we need to address the role of mobility, risk and health disparities on TB prevalence levels in a two patch setting. In the next section, we explore the role of certain parameters defining mobility, risk and health disparities, on the dynamics of TB  

\subsection{The role of risk and mobility on TB prevalence.}\label{sec:Scenarios}

We highlighted the dynamics of tuberculosis within a two patch system, described by (\ref{PatchGenFinal}), under various residence times schemes via numerical experiments. The simulations were carried out using the two-patch Lagrangian modeling framework on pre-constructed scenarios. We assume that one of the two regions (say, Patch 1) has high TB prevalence. We do not model specific cities or regions.

The interconnection of the two idealized patches demand that individuals from Patch 1  travel to the ``safer'' Patch 2 to work, to school or for other social activities. It is assumed that the proportion of time that Patch 2-residents spend in Patch 1 is negligible. 

In this study we define ``high risk'' based on the value of the  probability of developing active TB using two distinct definitions: in Section \ref{dt} high risk patch is defined by patch having high direct first time transmission potential but no difference in exogeneous reinfection potential between patches  ($\beta_1>\beta_2$ and $\delta_1=\delta_2$) and in Section \ref{er} high risk patch is determined by the patch with high exogenous reinfection potential ($\delta_1> \delta_2$ and $\beta_1=\beta_2$ ). In addition, we assume a fixed population size for Patch 1 and vary the population size of Patch 2. Particularly, we assume that Patch 1 is the denser patch while Patch 2 is assumed to have $\frac{1}{2}N_1$ and $\frac{1}{4}N_1$. That is, contact rates are higher in the Patch 1 population as compared to corresponding rates in Patch 2.

\subsubsection{The role of risk as defined by direct first time transmission rates}
\label{dt}

In this subsection, we explore the impact of differences in transmission rates between patches. Patch 1 is high risk  ($\mathcal{R}_{0}^1>1$; obtained by assuming $\beta_1 > \beta_2$) while Patch 2, in the absence of visitors would be unable to sustain an epidemic ($\mathcal{R}_{0}^2<1$). In addition the effect of different population ratios $\left( \dfrac{N_1}{N_2}\right)$ is explored.
 
\begin{figure}[H]
\centering
   \includegraphics[scale=0.4]{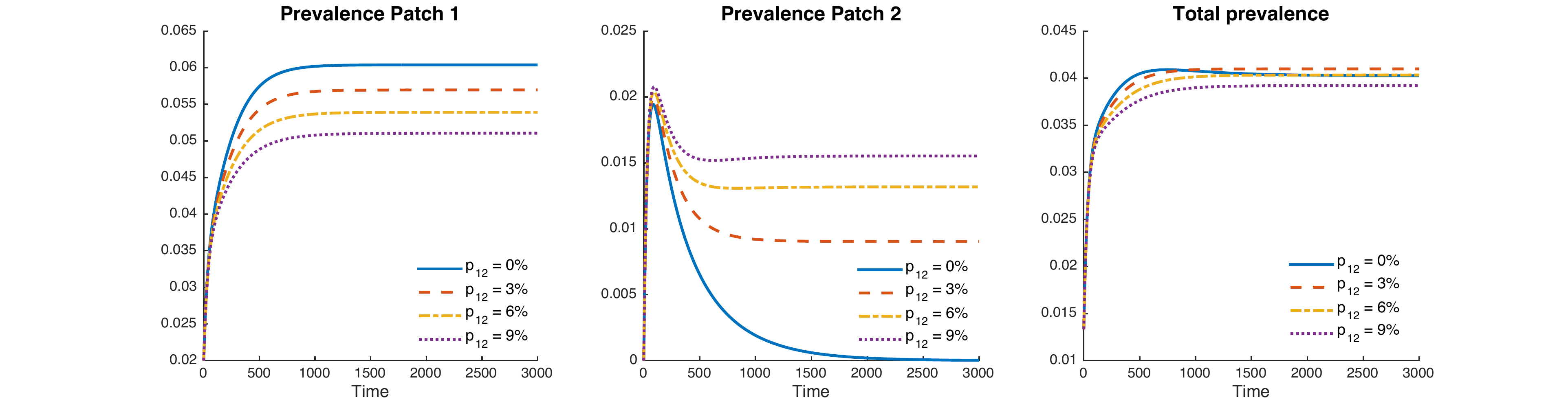}
\hspace{0.01\linewidth}%
   \includegraphics[scale=0.4]{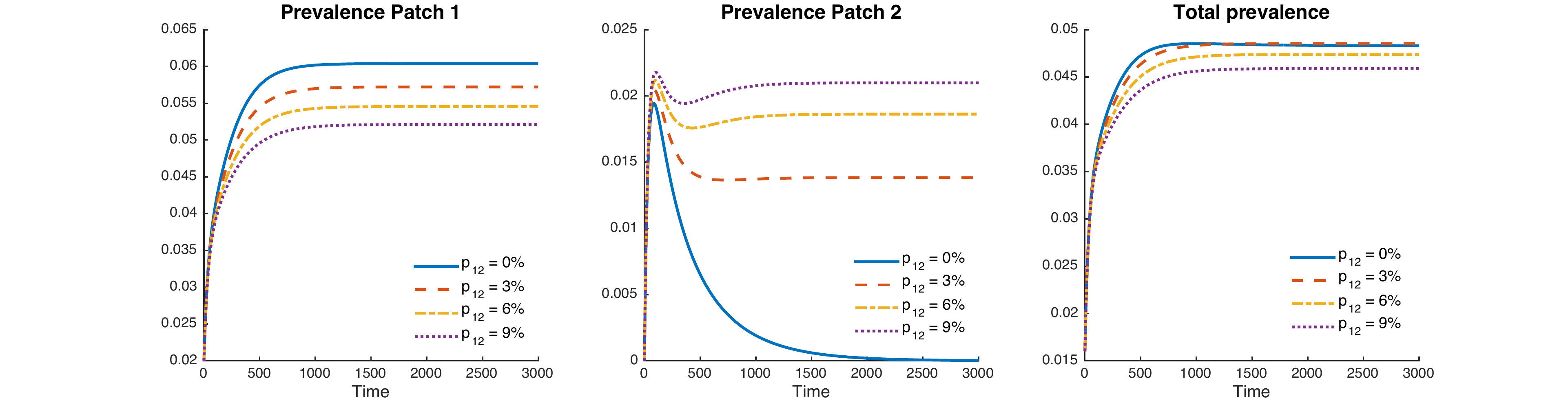}
\caption{Effect of mobility at $0\%$, $3\%$, $6\%$ and $9\%$ levels, for different transmission rates $0.13 = \beta_1 > \beta_2=0.07$ (which gives 
$\mathcal{R}_{0}^1=1.5$, $\mathcal{R}_{0}^2=0.8$) and $\delta_1=\delta_2=0.0026$, on the prevalence of TB over time. The cumulative prevalence and prevalence for each patch using the following population size proportions $N_2=\frac{1}{2}N_1$ (top figure) and $N_2=\frac{1}{4}N_1$ (bottom figure) are shown here.}
\label{prevtsrisk}
\end{figure}

Figure \ref{prevtsrisk} shows levels of patch prevalence reached when time-residency of Patch 1 individuals in Patch 2 is allowed at $0\%$ $3\%$, $6\%$ and $9\%$, within scenarios that assume that the population in Patch 1 is twice and four times that  of  Patch 2 and corresponding transmission values to match the assumed reproductive numbers in each patch. 

\textit{The results suggest that increments in mobility from Patch 1 to Patch 2 reduce TB prevalence in Patch 1 while increasing it in Patch 2. However, the number of total infected individuals from both patches slightly decreases, a global beneficial effect}.

\begin{figure}[H]
\centering
   \includegraphics[scale=0.4]{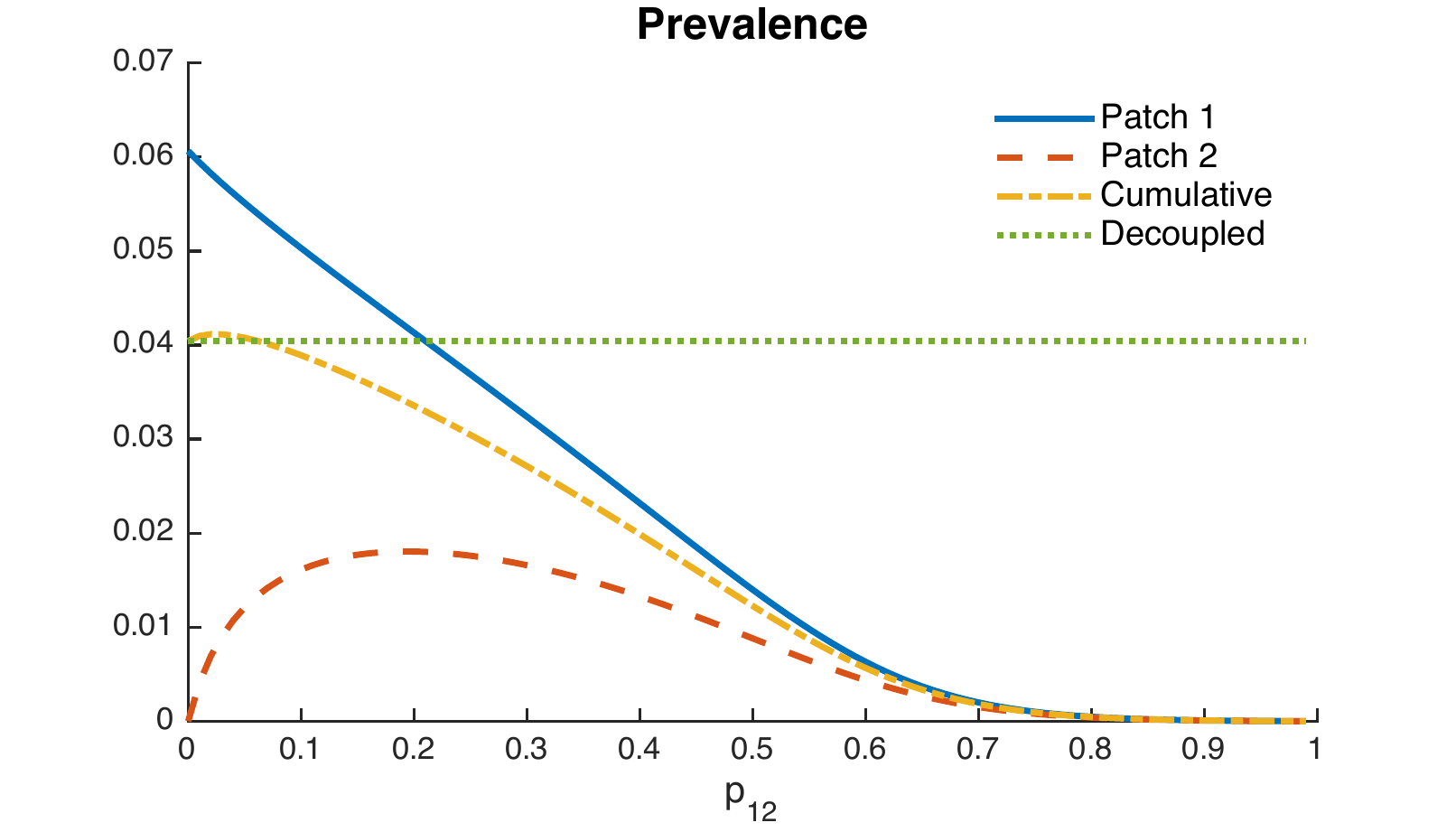}
\hspace{0.01\linewidth}%
   \includegraphics[scale=0.4]{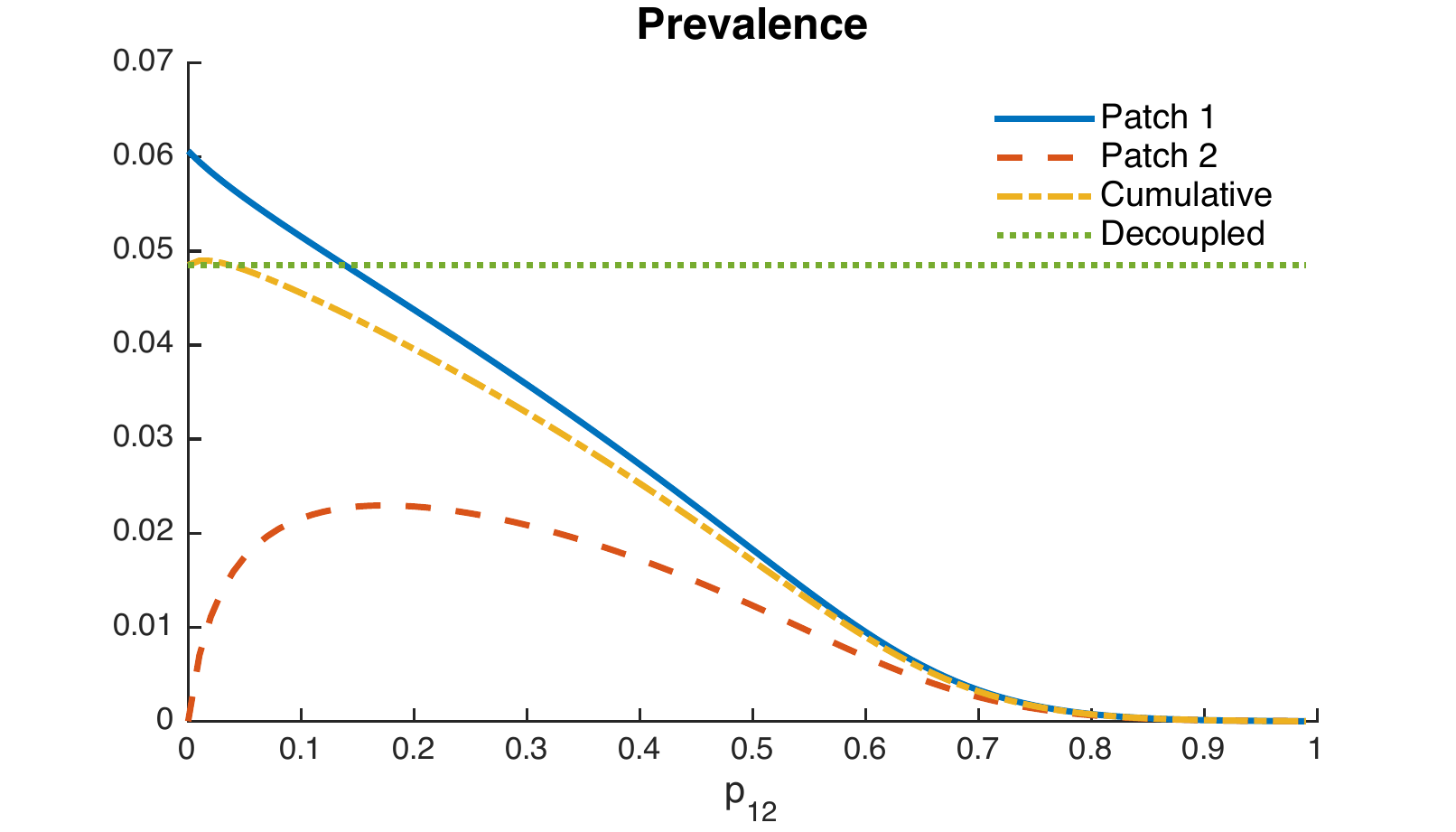}
\caption{Effect of mobility in the case of different transmission rates $0.13 = \beta_1 > \beta_2=0.07$ (which gives 
$\mathcal{R}_{0}^1=1.5$, $\mathcal{R}_{0}^2=0.8$) and $\delta_1=\delta_2=0.0026$, on the endemic prevalence. The cumulative prevalence and prevalence for each patch using the following population size proportions  $N_2=\frac{1}{2}N_1$ (left figure) and $N_2=\frac{1}{4}N_1$ (right figure) are shown here. The green horizontal doted line represents the decoupled case (i.e., the case when there is no movement between patches).}
\label{prevfnrisk}
\end{figure}

Figure \ref{prevfnrisk} uses mobility values $p_{12}$ as it looks at their impact on increases in cumulative two-patch prevalence. At the individual patch level, increase in mobility values reduce the prevalence in Patch 1 but increases the prevalence in Patch 2 initially and then decreases past a threshold value of $p_{12}$ (see Red and yellow curve in Fig 9a). That is, completely cordoning off infected regions may not be a good idea to control disease. However, movement rate of individuals between high risk infection region and low-risk region must be maintained above a critical value to control an outbreak.  

\textit{Thus, it is possible that when Patch 1 (riskier patch) has a bigger population size then mobility may turn out to be beneficial; the higher the ratio in population sizes, the higher the range of beneficial ``traveling'' times}. 
%
%
\subsubsection{The impact of risk as defined by exogenous reinfection rates}
\label{er}
Here, we focus our attention on the impact of exogenous reinfection on TB's transmission dynamics when transmission rates are the same in both patches, $\beta_1=\beta_2$. In this scenario, we assume the disease in both patches have reached an endemic state, that is, $\mathcal{R}_{0}^1>1$ and $\mathcal{R}_{0}^2>1$. However, Patch 1 remains the riskier, due to the assumption that exogenous reactivation of TB in Patch 1 is higher than in Patch 2, $\delta_1>\delta_2$.

\begin{figure}[H]
\centering
   \includegraphics[scale=0.4]{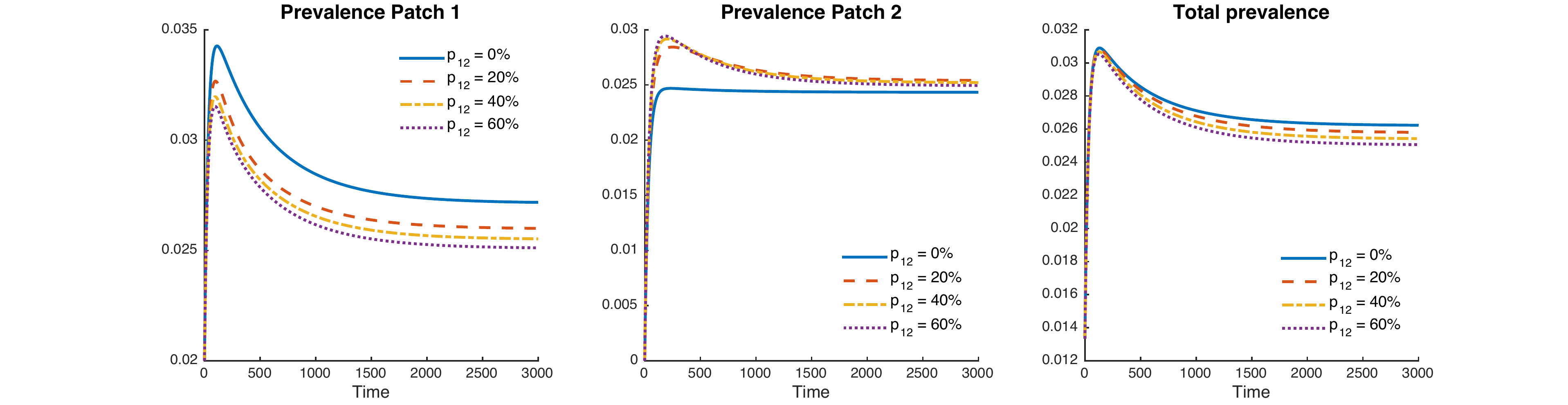}
\hspace{0.01\linewidth}%
   \includegraphics[scale=0.4]{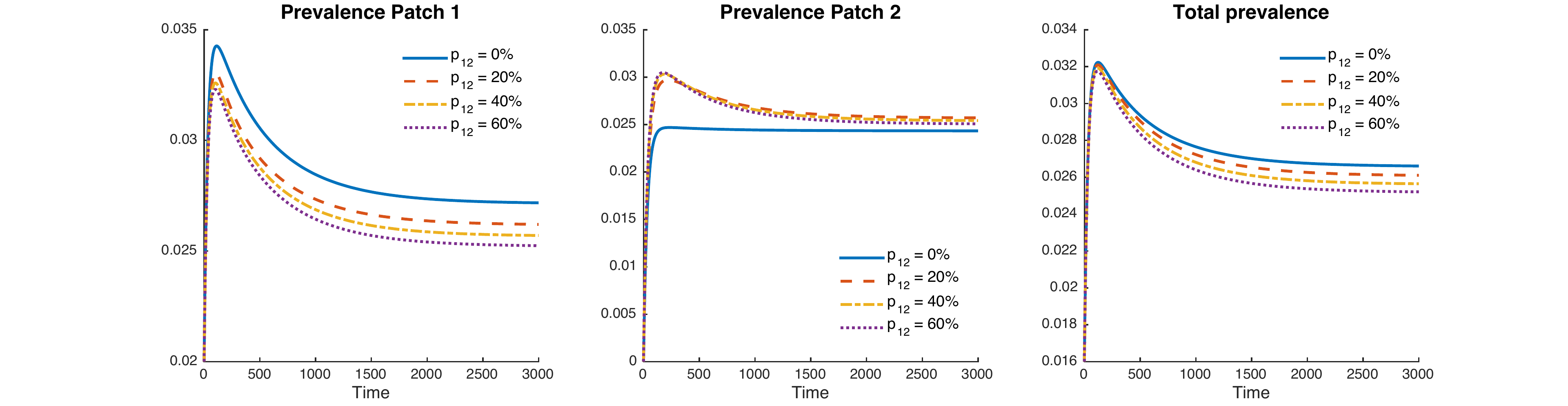}
\caption{Effect of mobility at $0\%$, $20\%$, $40\%$ and $60\%$ levels, when risk is defined by the exogenous reinfection rates $0.0053 = \delta_1 > \delta_2 = 0.0026$ and $\beta_1=\beta_2=0.1$ (which gives $\mathcal{R}_{0}^1=\mathcal{R}_{0}^2=1.155$), on the prevalence over time. The cumulative prevalence and prevalence for each patch using the following population size proportions $N_2=\frac{1}{2}N_1$ (top figure) and $N_2=\frac{1}{4}N_1$ (bottom figure) are shown here.}
\label{prevtsexog}
\end{figure}

Figure \ref{prevtsexog} shows levels of patch prevalence when individuals from Patch 1 travel to Patch 2, $0\%$ $20\%$, $40\%$ and $60\%$ of their time. The cases where the population in Patch 1 is twice and four times the size of the population in Patch 2 are explored. As in the previous case, prevalence levels in Patch 1 are being reduced by mobility, while prevalence is being increased in Patch 2.

\begin{figure}[H]
\centering
\includegraphics[scale=0.4]{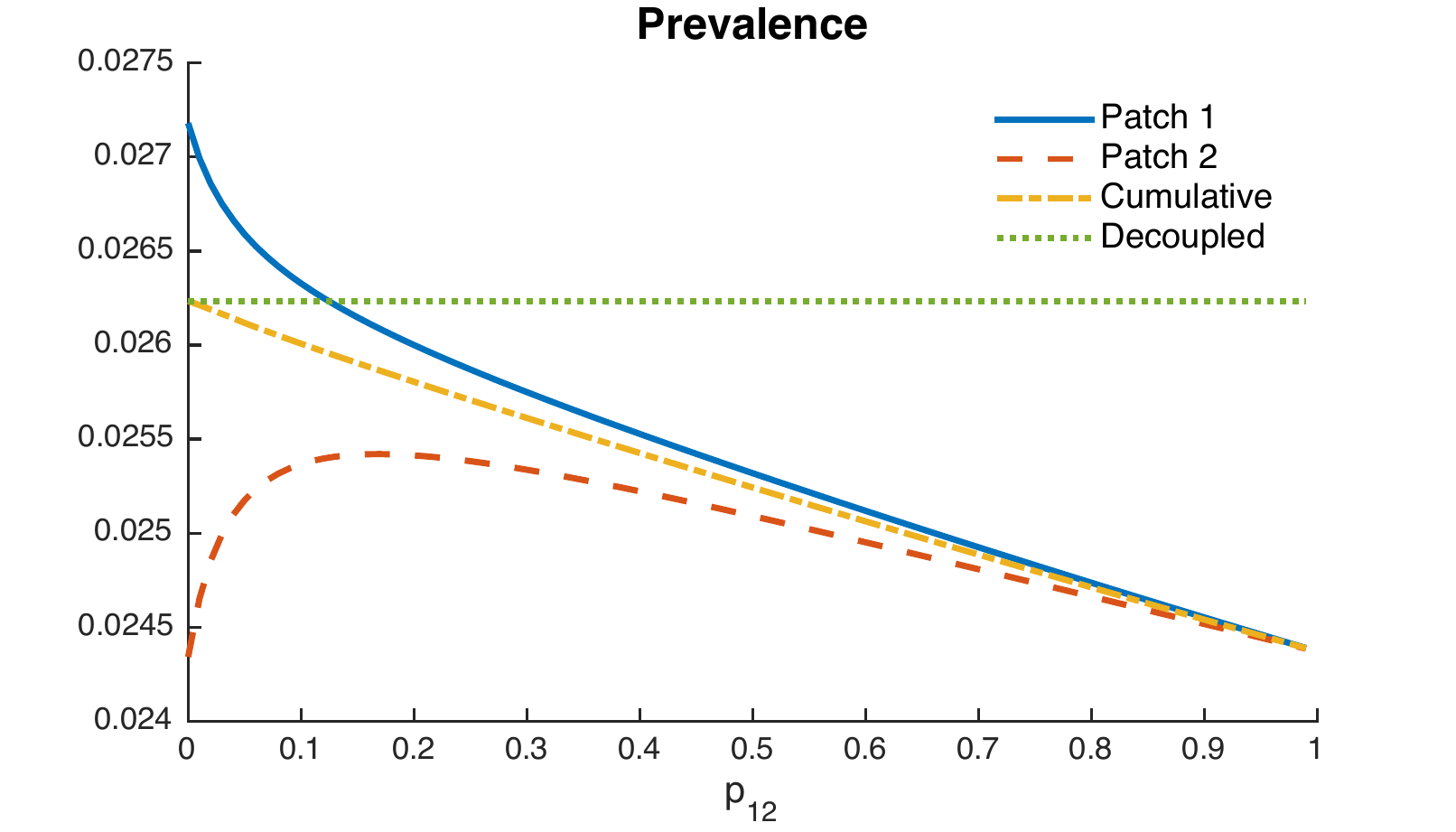}
\hspace{0.01\linewidth}%
   \includegraphics[scale=0.4]{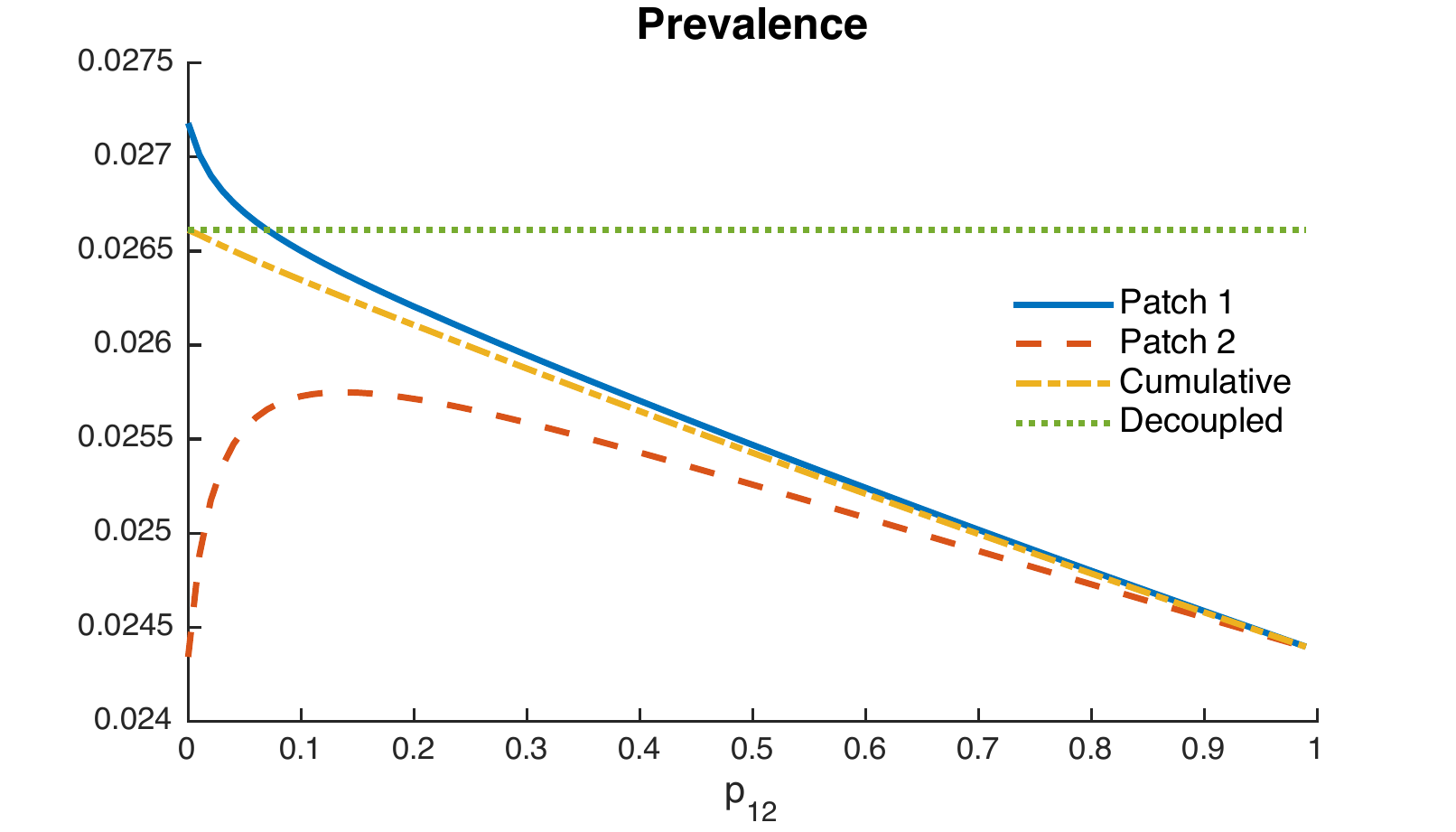}
\caption{Effect of mobility when risk is defined by the exogenous reinfection rates $0.0053 = \delta_1 > \delta_2 = 0.0026$ and $\beta_1=\beta_2=0.1$ (which gives $\mathcal{R}_{0}^1=\mathcal{R}_{0}^2=1.155$), on the endemic prevalence. The cumulative prevalence and prevalence for each patch using the following population size proportions $N_2=\frac{1}{2}N_1$ (left figure) and $N_2=\frac{1}{4}N_1$ (right figure) are shown here. The green doted line represents the decoupled case  (i.e., the case when there is no movement between patches).}
\label{prevfnexog}
\end{figure}

Nevertheless, the reduction of prevalence in Patch 1 is bigger than the increment in Patch 1; thus, the effect on the overall system is represented by a reduction of prevalence for the explored values of $p_{12}$.
Figure \ref{prevfnexog} shows the combined role of exogenous reinfection and mobility values when the population of Patch 1 is twice or four times the population of Patch 2. 

It is possible to see a small reduction in the overall prevalence, given for all mobility values from Patch 1 to Patch 2.

\textit{Within this framework, parameters and scenarios, our model suggest that direct first time transmission plays a central role on TB dynamics when mobility is considered. Although mobility also reduces the overall prevalence when exogenous reinfection differs between patches, its impact is small as compared to direct first time transmission results}.

\section{Discussion}\label{sec:Discussion}
According to the world health organization (W.H.O) \cite{WHO2015}, in 2014, 80\% of the reported TB cases occurred in 22 countries, all developing countries. Efforts to control TB have been successful in many regions of the globe and yet, we still see 1.5 million people die each year. And so, TB, faithful to its history \cite{Daniel2006}, still poses one of the greatest challenges to global health. Recent reports suggest that established control measures for TB have not been adequately implemented, particularly in sub-Saharan countries \cite{Andrews2013,Chatterjee2015}. In Brazil rates has decreased but relapse is more important than reinfection \cite{Oliveira2013,Luzze2013}. Finally, in Cape Town, South Africa, a study \cite{Verver2005} showed that in high incidence areas, individuals who have received TB treatment and are no longer infectious are at the highest risk of developing TB instead of being the most protected.

Hence, policies that do not account for population specific factors are unlikely to be effective. Without a complete description of the attributes of the community in question, it is almost impossible to implement successful intervention programs that are capable of generating low reinfection rates through multiple pathways and low number of drug resistant cases. Intervention programs must educate populations and their government officials on the benefits, factors, and cost associated with population-based TB prevention and control programs. Intervention must account for the risks that are inherent with high levels of migration as well as with local and regional mobility patterns between areas defined by high differences in TB risk. 

In this manuscript, we have focused on the role of `daily' mobility within high and low-risk areas and their potential impact on TB dynamics and control. A situation that is not so uncommon in areas where extreme levels of social, economic and health disparities rule. We carry out the discussion using a simplified framework, that is, a two-patch system, that captures, in a rather `dramatic' way the dynamics between two worlds; the world of the haves and the have nots. The results are highlighted via the simulation of simplified extreme scenarios as the main objective of this manuscript is to stress the impact of disparities.

The model analysis suggests that dynamics of TB depends on the basic reproduction number ($\mathcal{R}_0$), which in turn is the function of model parameters that includes direct first transmission and exogenous (reinfection) transmission rates. The simulations of specific extreme scenarios suggest that short term mobility between heterogeneous patches does not always contributes to overall increases in TB prevalence. The results show that when risk is considered only in terms of exogenous reinfection, the global TB prevalence remains almost unchanged, compared to the effect of direct new infection transmission. In the case of a high risk direct first time transmission, it is observed that mobile populations may pose detrimental effects on the prevalence levels in both environments (patches). The simulation show that when the individuals from the risky population spend  $25\%$ of their time or less in the safer patch is bad for the overall prevalence. However, if they spend more, the overall prevalence decreases. Further, in the absence of exogenous reinfections, the model is robust, that is, the disease dies out or persists based on whether or not the basic $\mathcal{R}_0$ is below or above unity, respectively. Although, the role of exogenous reinfection seems not that relevant on overall prevalence, the fact remains that such mode of transmission increases the risk that come from large displacement of individuals, due to catastrophes or conflict, to TB-free areas.

Our ability to interpret information regarding the local origin of mobile individuals accurately would facilitate prompt responses in the face of initiation of an epidemic. During the development and implementation of training and educational programs the necessity to {\it avoid stigmatizing and further marginalization of groups that may have already experienced some kind of discrimination} is essential to avoid isolation, prevent integration, and reduce compliance \cite{Gushulak2000}. A situation that cannot be ignored in today's world where conflicts have dislocated the lives of millions and generated new migration patterns that includes millions of refuges.

Failure to adequately incorporate and address these challenges may result in considerable delays. As noted in \cite{Feng2000}, ignoring exogenous reinfections, that is, establishing policies that focus exclusively on the reproductive number $\mathcal{R}_0$, would amount to ignoring the role of dramatic changes in initial conditions, now more common than before, due to the displacement of large groups of individuals, the result of catastrophes and conflict.\\

%
\textbf{Acknowledgements:}
{\small{This project has been partially supported by grants from the National Science Foundation (NSF - Grant DMPS-0838705), the Alfred P. Sloan Foundation, the Office of the Provost of Arizona State University and the National Institute of General Medical Sciences (NIGMS) at the National Institutes of Health (NIH - Grant \#1R01GM100471-01). The contents of this manuscript are solely the responsibility of the authors and do not necessarily represent the official views of DHS or NIGMS. The funders had no role in study design, data collection and analysis, decision to publish, or preparation of the manuscript.
}}
\bibliographystyle{elsarticle-num}


\end{document}